\documentclass{aa}
\usepackage{graphics}
\newcommand{\CC}{C$_{\rm 2~}$}
\newcommand{\fwm}{$\Delta_{g}$}
\newcommand{\Tef}{$T_{\rm eff}$}

\newcommand{\EBV}{$E_{\rm B-V}$}
\begin{document}

\thesaurus{08(08.01.1; 08.16.4; 08.09.2: Sakurai's object)}

\title{Modeling the spectrum of V4334 Sgr (Sakurai's object)}

\author{Ya.V. Pavlenko \inst{1} \and L.A. Yakovina\inst{1} 
\and H.W. Duerbeck \inst{2}}

\institute{Main Astronomical Observatory of Academy of Sciences of
Ukraine, Golosiiv woods, Kyiv-127, 03680 Ukraine\\
e-mail: yp@mao.kiev.ua
\and
WE/OBSS, Free University Brussels (VUB), 
Pleinlaan 2, B-1050 Brussels, Belgium \\
e-mail: hduerbec@vub.ac.be}

\offprints{Ya.V. Pavlenko}

\date{Received date / Accepted date}

\maketitle

\titlerunning{Modeling the spectrum of V4334 Sgr}

\begin{abstract}
Theoretical spectral energy distributions were computed for a grid of 
hydrogen-deficient and carbon-rich model atmospheres of \Tef{} in
the range of $5000-6250$~K and $\log\, g = 1.0 - 0.0$ by the technique 
of opacity sampling, taking into account continuous, molecular band 
and atomic line absorption. These energy distributions were compared 
with the spectrum of V4334 Sgr (Sakurai's object) of April, 
1997 in the wavelength interval $300-1000$ nm. We show that
(1) the shape of the theoretical spectra depends strongly on \Tef{},
but only very weakly on the hydrogen abundance;
(2) the comparison of the observed and computed spectra permits to
estimate \Tef{} $\approx 5500$ K for V4334 Sgr in April, 1997, and
its interstellar reddening (plus a possible circumstellar
contribution) $E_{\rm B-V}\approx 0.70$.


\end{abstract}

\section{Introduction}

\object{V4334 Sgr} (Sakurai's object) was discovered on February 20, 
1996 as a ``novalike object in Sagittarius'' at magnitude $\sim 11^{\rm m}$ 
(Nakano et al. 1996). At the time of discovery, it showed an absorption 
line spectrum of type F with unusually strong lines of C~{\sc{i}}, 
C~{\sc{ii}} and O~{\sc{i}}. Its progenitor was a faint blue star of 
magnitude $\sim 21^{\rm m}$ in the center of an uncatalogued planetary 
nebula of low surface brightness.

Prediscovery photographs showed that V4334 Sgr had increased in optical 
luminosity already during 1995. Possibly, it was an UV-bright object in 
the beginning of the outburst, and kept a almost constant bolometric 
luminosity during most of its evolution. First indications of the presence 
of dust were seen as a possible IR excess in 1996 (Duerbeck \& Benetti 1996). 
This excess increased in strength in the following years. In early and in 
late 1998, abrupt brightness declines occurred at optical wavelengths, 
which can be interpreted as more and more dramatic events of dust formation, 
which have some resemblance to those of R CrB variables (Duerbeck et al. 
1999a,b), as well as to those in the post-AGB object \object{FG Sge} 
(Gonzalez et al. 1998). 

The nature of the photometric and spectroscopic evolution indicates 
that V4334 Sgr is undergoing a final helium flash (Duerbeck \& Benetti 1996). 
Such events have been investigated theoretically in connection with 
the slowly evolving final helium flash object FG Sge (e.g. Fujimoto 1977, 
Iben et al. 1983).

The exact spectral classification of V4334 Sgr is complicated because of the 
unusual abundances. Its atmosphere shows a hydrogen deficit, as well as 
C/O $>$ 1, and $^{12}$C/$^{13}$C $\sim$ 2.5 (Asplund et al. 1997, Kipper 
\& Klochkova 1997). The abundances appear to change on short time scales. 
These changes are restricted to decreasing H, increasing Li and s-process 
elements, and possibly increasing Ti and Cr (Asplund et al. 1999).

In this paper, we study some problems related to the modeling of the spectral 
energy distribution (SED) of V4334 Sgr. This requires computations 
of model atmospheres with peculiar chemical abundances and theoretical 
spectra which consist of numerous atomic lines and molecular bands.

\section{Procedure}

\subsection{Observational material}

A flux-calibrated spectrum of V4334 Sgr in the range $\lambda\lambda$ 
300-1000 nm, taken April 29, 1997 with the ESO 1.52\,m telescope at La 
Silla (see Duerbeck et al. 1997, 1999a for more details), corrected for 
interstellar extinction, is used for comparison with theoretical spectra.

The interstellar extinction is still an unclear point. Duerbeck \& Benetti 
estimated \EBV{} = 0.54 from the Balmer decrement of the lines of the 
planetary nebula, and Duerbeck et al. (1997) obtained 0.53 from a comparison 
with synthetic colors of hydrogen-deficient stars, Eyres et al. (1998) derive 
1.15 from the observed versus expected H$\beta$ flux. Pollacco (1999), 
using high S/N spectra of the planetary nebula, derived $0.71 \pm 0.09$ 
from the Balmer decrement. Kimeswenger \& Kerber (1998) determined the 
interstellar extinction as a function of distance, and arrive at a value
$\approx 0.8$ for distances above 1 kpc. We assume in this paper extinction 
values \EBV{} = 0.53 and 0.7, and check which one yields better agreement 
between computed SEDs and observations, dereddened for the above values. 
In 1998 and 1999, increasing amounts of circumstellar dust also modified 
the observed SED of the central object (Kipper 1999). At the time of our 
observation, the influence of circumstellar extinction appears to be 
marginal (see also section 4).

\subsection{Model atmospheres}

Opacity sampling model atmospheres of different \Tef{}, $\log\, g$ and 
abundances are computed with the program SAM941 (Pavlenko 1999). Chemical 
abundances derived by Asplund et al. (1997) are used as ``normal input'' 
for V4334 Sgr. Asplund et al. (1999) derived abundances for October 1996, 
which are probably best to be adopted in the absence of estimates for 
subsequent dates. Nevertheless, we do not expect our results to be 
crucially affected by abundances which differ from those of Asplund et al. 
(1997). We varied a few abundances to study the impact of abundance 
changes on the emitted spectrum. Results are given in section 3.4. 

The ionization-dissociation equilibrium (IDE) was calculated for a mix of 
70 atoms, ions and diatomic molecules. Constants for IDE computations were 
taken mainly from Tsuji (1973). Absorption by atoms and ions as well as 
absorption in frequencies of 20 band systems of diatomic molecules were 
taken into account (Table 1).

\begin{table}
\caption{Band systems of the absorbing two-atomic molecules}
\begin{tabular}{llrrl}\hline\hline
\multicolumn{2}{l}{molecule system} & 
$\lambda_1$~\phantom{.} & $\lambda_2$~\phantom{.} & \\ \hline
 \CC & e$^3\Pi_g$-a$^3\Pi_u$    &  237.0 &  328.5  & Fox-Herzberg  \\
 \CC & A$^1\Pi_u$-X$^1\Sigma^{+}_g$  & 672.0 & 1549.0 & Phillips      \\
 \CC & b$^3\Sigma^-_g$-a$^3\Pi_u$  & 1100.0 & 2700.0  & Ballik-Ramsay \\
 \CC & d$^3\Pi_g$-a$^3\Pi_u$  &  340.0 &  785.0  & Swan \\
 CN  & A$^2\Pi$-X$^2\Sigma^+$ &  400.0 & 1500.0  & red     \\
 CN  & B$^2\Sigma^+$-X$^2\Sigma^+$  & 240.0 &  600.0  & blue       \\
 CS  & A$^1\Pi$-X$^1\Sigma^+$  &  240.0 &  330.0  & \\
 CO  & B$^1\Sigma^+$-A$^1\Pi$  &  412.0 &  662.0  & \AA{}ngstr\"{o}m \\
 CO  & C$^1\Sigma^+$-A$^1\Pi$  &  368.0 &  571.0  & Herzberg   \\
 CO  & A$^1\Pi$-X$^1\Sigma^+$  &  114.0 &  280.0  & \\
 NO  & C$^2\Pi_r$-X$^2\Pi_r$   &  207.0 &  275.0  & $\delta$ \\
 NO  & B$^2\Pi_r$-X$^2\Pi_r$    & 200.0 &  380.0  & \\
 NO  & A$^2\Sigma^+$-X$^2\Pi_r$  & 195.0 &  340.0  & $\gamma$  \\
 MgO & B$^1\Sigma^+$-X$^1\Sigma^+$  & 454.0 &  544.0  &     \\
 AlO & C$^2\Pi_r$ -X$^2\Sigma^+$     & 200.0 &  400.0  &   \\
 AlO & B$^2\Sigma^+$-X$^2\Sigma^+$  & 404.0 &  580.0 &  \\
 SiO & E$^1\Sigma^+$-X$^1\Sigma^+$  & 171.5 &  200.0  & \\
 SiO & A$^1\Pi$-X$^1\Sigma^+$   &  207.0 &  330.0  & \\
 SO  & A$^3\Pi$-X$^3\Sigma^-$   &  246.0 &  380.0  & \\
 CaO & C$^1\Sigma$-X$^1\Sigma$ &   730.0 &  923.0  & \\ \hline\hline
\end{tabular}

\noindent  $\lambda_1$, $\lambda_2$ (nm) -- the range of wavelengths in which
molecular band absorption was taken into account. 
\end{table}

The atomic line list was taken from the VALD database (Piskunov etal. 1995), 
including lines of s-process elements which are present in the spectrum 
of V4334 Sgr. Molecular opacities were computed in the approach of the just 
overlapping approximation (JOLA). The JOLA approach is based on the assumption
that the mean line separation within a molecular band is comparable or 
smaller than the line widths. By definition, JOLA overestimates molecular 
absorption produced by weak molecular bands. For strong (saturated) 
molecular bands, the results appear to be satisfactory (Pavlenko 1997). 
We used the BIGF1 program of Yaremchuk (Nersisyan et al. 1987) which 
realizes the method of Kamenshikov et al. (1971). Yaremchuk's approach 
takes into account the splitting of molecular bands on the P-Q-R or P-R 
branches for systems with $\Lambda =1, 0$, respectively. The computations 
were carried out for a set of vibrational quantum numbers $0 \leq v^{\prime}, 
v^{\prime\prime} \leq 9 $ (see Abia et al. 1999 and Pavlenko \& Yakovina 
1999 for more details). 

Convection was computed using the ATLAS9 (Kurucz 1993) scheme with
$l/H = 1.6$. Convective overshooting was not considered. Our numerical 
experiments show that the impact of convective overshooting on the 
temperature structure of the model atmospheres is rather weak.

The main opacity sources in the atmosphere of V4334 Sgr differ from those in 
atmospheres with solar-like abundances:
\begin{itemize}
\item bound-free transitions of C~{\sc{i}}  for higher temperatures
(as in R CrB star atmospheres). We used Hoffsaess' (1979) tables of 
bound-free opacities due to C~{\sc{i}} absorption. The comparison with 
data from the Opacity Project (Seaton et al. 1992) shows good 
agreement. More details of the continuous opacity computation 
procedure are given in Pavlenko (1999); 
\item bound-bound transitions of atoms and ions (line haze) in the 
near-UV spectrum;
\item electronic transitions of diatomic molecules (molecular bands) 
for lower temperatures.
\end{itemize}

To take into account both non-solar abundances and opacity sources, 
we computed tables of Rosseland opacities $\tau_{\rm ross}$ for a grid 
of temperatures $T$ and pressures $P$. The table $\tau_{\rm ross} = 
f (T,P)$ was used for the temperature correction in SAM941. We used the 
same grid of opacity sources as in model atmosphere computations, i.e.
continuous, molecular band and atomic line absorptions. This procedure 
is important since the photosphere of V4334 Sgr lies at a different pressure 
height that in the case of solar abundances (cf. the computations 
for R CrB, Pavlenko 1999). 

\begin{figure}
\resizebox{\hsize}{!}{\includegraphics{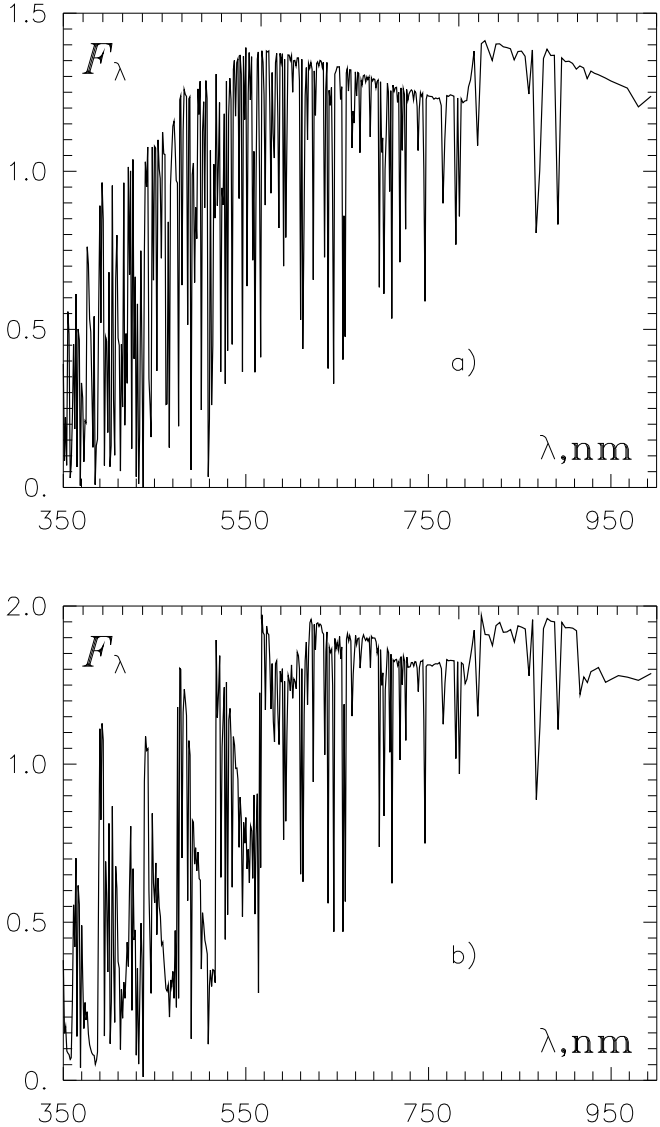}}
\caption{Theoretical energy distributions computed for model atmospheres 
6250/1.0 of V4334 Sgr (a) without and (b) with molecular (JOLA) absorption.}
\label{fig1}
\end{figure}

\begin{figure}
\resizebox{\hsize}{!}{\includegraphics{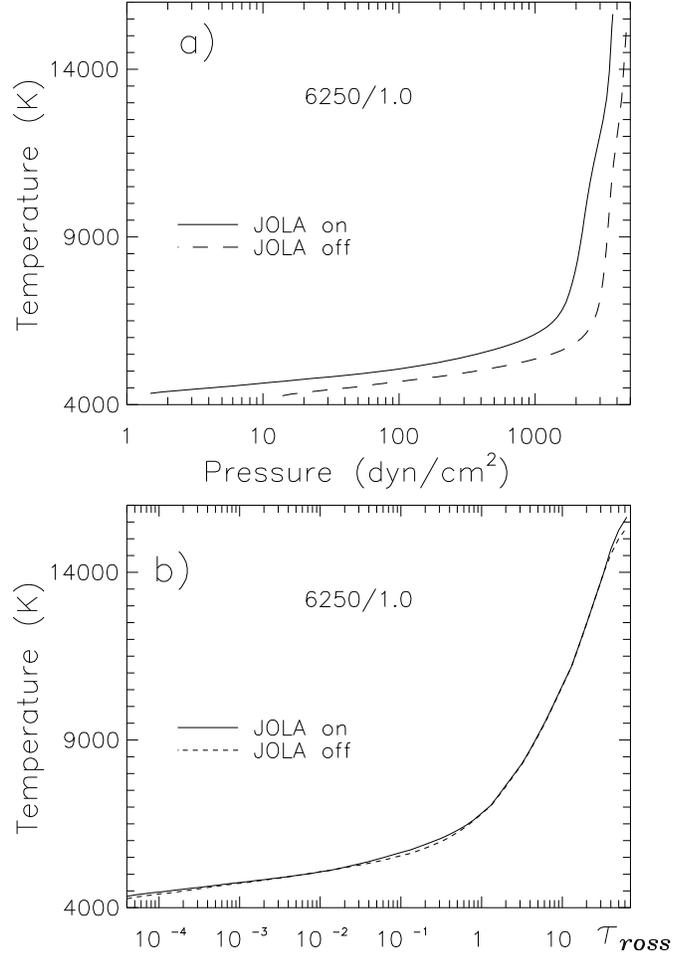}}
\caption{The impact of the molecular (JOLA) absorption on the temperature 
structure of the 6250/1.0 model atmosphere of V4334 Sgr shown as (a) $T$ vs. 
$P_g$ and (b) $T$ vs. $\tau_{\rm ross}$.}
\label{fig2}
\end{figure}

Strong molecular bands clearly appear in spectra computed for model 
atmospheres with \Tef{}$/\log\, g = 6250/1.0$  (Fig.~\ref{fig1}). As 
a consequence of the appearance of additional molecular absorption, 
the temperature distribution $T=f(\tau_{\rm ross})$ in the atmosphere 
is shifted to lower pressures even in the case of such a comparatively 
high \Tef{} (Fig.~\ref{fig2}a). This is because an increase of opacity 
in the outer part of the atmosphere changes its temperature structure, 
and even the temperatures in the (sub)photospheric layers ``respond'' 
to these changes. The changes of model atmospheres in the scale 
$\tau_{\rm ross}$ are less pronounced (Fig.~\ref{fig2}b).

Molecular bands become stronger when \Tef{} drops below 6000 K, because
the densities of CN and \CC molecules, which are the main absorbers,
increase with decreasing \Tef{}.


\subsection{Algorithm}

Numerical computations of theoretical radiative fluxes $F_{\lambda}$ 
in the spectrum of V4334 Sgr were carried out within the classical approach: 
LTE, plane-parallel model atmosphere, no energy divergence.

We studied the spectrum formation within a self-consistent approach, 
i.e. model atmospheres and theoretical SEDs are computed for the same 
opacity source list and sets of abundances. This approach allows us to 
treat the possible impact of abundance changes on the temperature 
structure of the model atmosphere as well as on the emitted fluxes 
in a direct way.

We modeled the SED emitted by V4334 Sgr in the region 
from the near UV to the far red\footnote{For simplification, we will use 
for ``theoretical energy distributions in the spectrum of V4334 Sgr'' 
the definition ``theoretical spectra''}. Computations were carried out 
with wavelength steps $\Delta\lambda$ = 0.1 nm and 0.005 nm. The 
theoretical spectra were then convolved by a Gaussian with half-width 
\fwm{} = 0.5 nm. A comparison of theoretical spectra computed with 
different $\Delta\lambda$ with the observed spectrum yields practically 
the same results.

To compare observed and theoretical spectra of V4334 Sgr, we normalize them, 
i.e. we adopt equal fluxes at $\lambda = 570$ nm. A choice of any 
other point for the normalization would not change our main conclusions.

\section{Results}

\subsection{Identification of the main absorbing features in the
spectrum of V4334 Sgr}

\begin{figure}
\resizebox{\hsize}{!}{\includegraphics{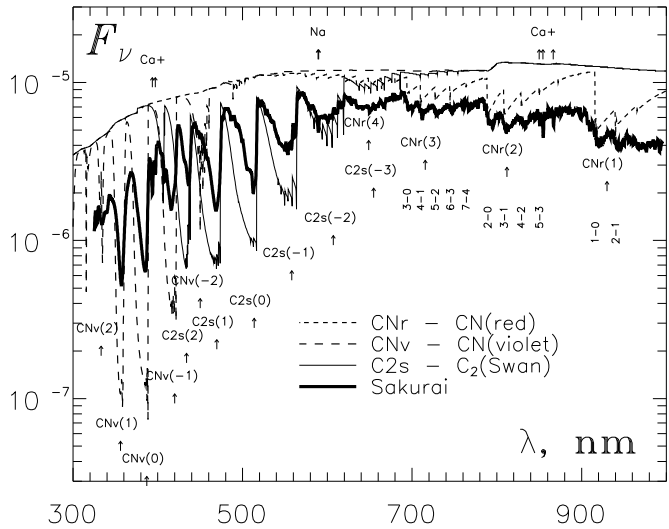}}
\caption{Identification of the strongest molecular and atomic features in 
the spectrum of V4334 Sgr in April, 1997. For the \CC bands, the difference of 
the vibrational quantum numbers $\Delta v = v^{\prime} - v^{\prime\prime}$ 
is shown in brackets. For the CN bands, pairs of $v^{\prime} - 
v^{\prime\prime}$ are shown.}
\label{fig3}
\end{figure}

Several molecular features in the April, 1997 spectrum of V4334 Sgr are 
identified in Fig.~\ref{fig3}. Its overall shape in the blue and
red parts of the spectrum is governed by the \CC and CN band systems,
respectively. In the near UV ($\lambda < 400~\rm nm$), atomic 
absorption becomes important. In the low-resolution spectrum, only 
the strongest atomic lines can be identified: Na {\sc i} D (589.16, 
589.75 nm), Ca {\sc ii} H and K (393.48, 396.96 nm), and the 
Ca {\sc ii}  IR triplet (850.03, 854.44, 866.45 nm).

\subsection{ The dependence on \Tef, and the temperature of V4334 Sgr in 
April, 1997}

\begin{figure}
\resizebox{\hsize}{!}{\includegraphics{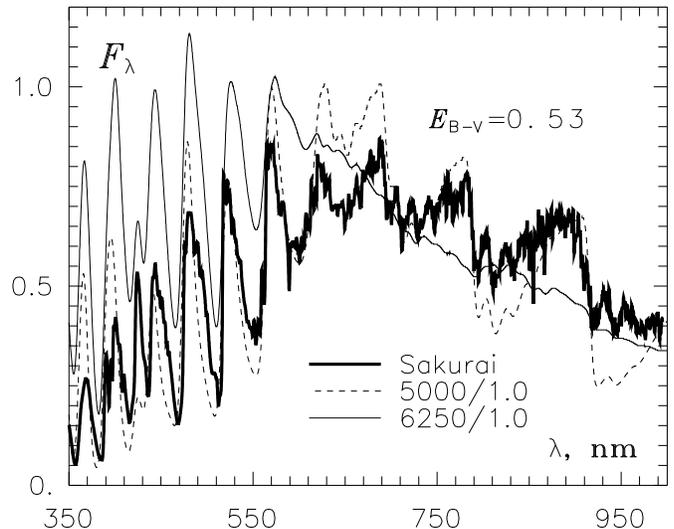}}
\caption{Comparison of the computed radiative field fluxes for model 
atmospheres with \Tef{} = 5000 and 6250 K and $\log\, g = 1.0$ with 
the observed spectrum of V4334 Sgr, corrected for \EBV{} = 0.53.}
\label{fig4}
\end{figure}

We computed theoretical SEDs for model atmospheres with \Tef{} = 5000, 
5250, 5500, 5750, 6000 and 6250~K. Fig.~\ref{fig4} shows the comparison 
of observed and computed fluxes $F_{\lambda}$ for model atmospheres 
5000/1.0 and 6250/1.0. The overall shape of the SED shows a critical 
dependence on \Tef{}. The comparison of observed and computed SEDs shows a 
reasonably good {\em qualitative\/} agreement. Thus, the contribution of 
various molecular opacity sources in the atmosphere of V4334 Sgr can be 
considered as well established.

\begin{figure}
\resizebox{\hsize}{!}{\includegraphics{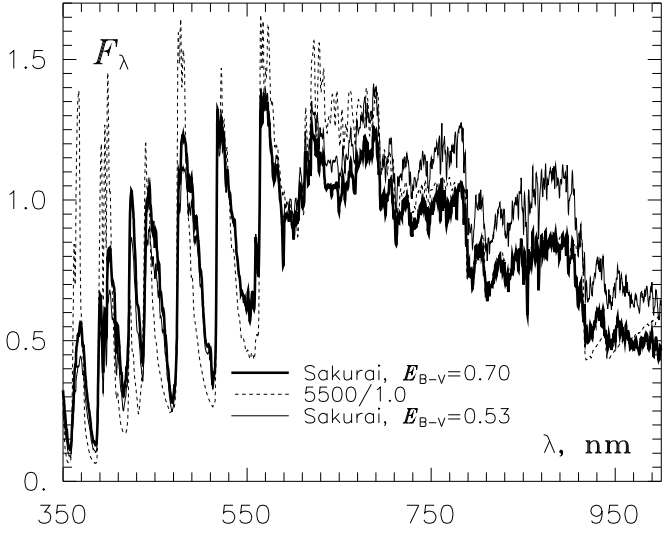}}
\caption{Comparison of the computed radiative field fluxes for model 
atmospheres with \Tef{} = 5500 K and the observed spectrum of V4334 Sgr, 
corrected for \EBV{} = 0.53 and 0.70.}
\label{fig5}
\end{figure}

Better fits can be obtained for the intermediate \Tef{} of 5500 K
(Fig.~\ref{fig5}). We estimate that the formal uncertainties of our 
\Tef{} determination do not exceed 200 K (see, however, section 4). 
In Fig.~\ref{fig5}, our theoretical spectra are compared with observed 
ones, corrected for different values of \EBV{}. We find that the fit with 
\EBV{} = 0.70 gives a better result. An especially good agreement is 
achieved in the red part of the spectrum, where the CN bands dominate.
In the UV and blue regions where molecular band opacity is minimal, the 
peaks in the computed flux distributions are too high in comparison with 
observations. This is most probably the consequence of a ``missing 
opacity'' at these wavelengths (see also section 4).  This ``missing 
opacity'' vanishes towards the red.

\subsection{The dependence on $\log\, g$}

\begin{figure}
\resizebox{\hsize}{!}{\includegraphics{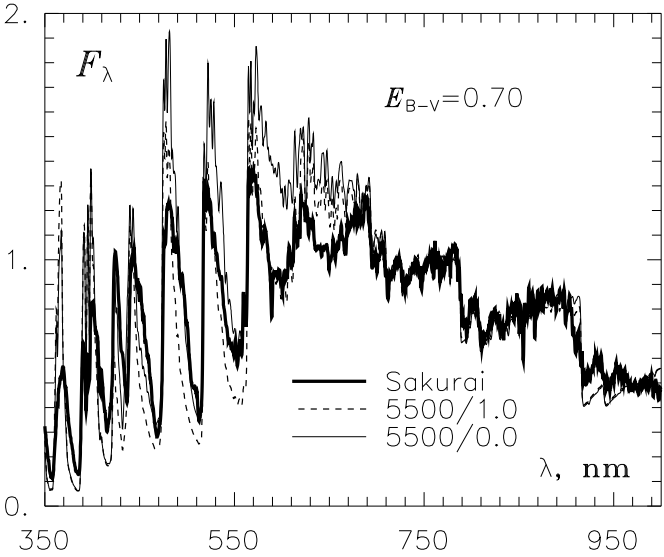}}
\caption{Comparison of the computed radiative field fluxes for model 
atmospheres 5500/1.0 and 5500/0.0 with the observed spectrum of V4334 Sgr, 
corrected for \EBV = 0.70.}
\label{fig6}\end{figure}

To study the impact of $\log\, g$ on the spectrum of V4334 Sgr, two model 
atmospheres 5500/1.0 and 5500/0.0 and their theoretical SEDs 
were computed. Results are shown in Fig.~\ref{fig6}. We note:
\begin{itemize}
\item the spectrum shape is much more dependent on \Tef{} than on 
$\log\, g$; the latter is of second order importance. In the red part 
of the spectrum, the impact of a change in $\log\, g$ is rather weak;
\item a moderately strong dependence of the spectrum shape on $\log\, g$
is found at $\lambda\lambda$ 570-620 nm; this region may be used for 
the determination of $\log\, g$;
\item comparing the observed spectrum of V4334 Sgr in April, 1997 with 
the computed spectra, we find that $\log\, g$ lies in the range 
$0 < \log\, g < 1$.
\end{itemize}

\subsection{The dependence on chemical abundances}

The chemical abundances in the atmosphere of V4334 Sgr changed on short time 
scales (see Asplund et al. 1997 for details). On the other hand, only 
for 1996 has the chemical composition of Sakurai's object been shown 
to vary (although it is not a bad guess that it continued to vary also 
in 1997).

\begin{figure}
\resizebox{\hsize}{!}{\includegraphics{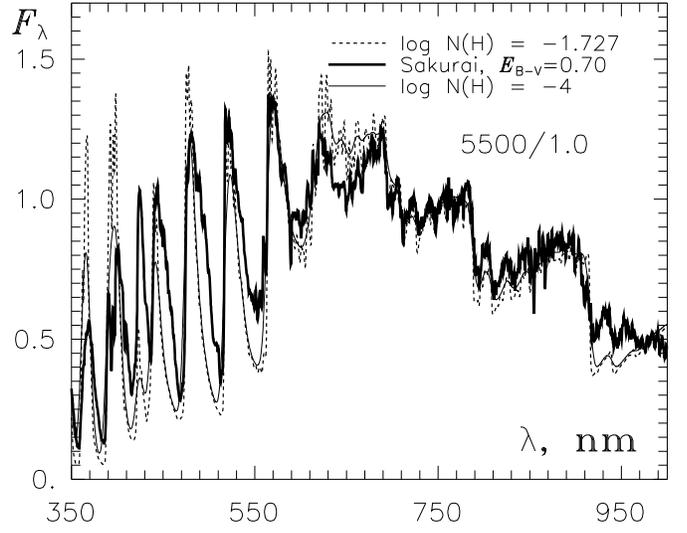}}
\caption{The dependence of theoretical spectra of V4334 Sgr on the
input hydrogen abundance.}
\label{fig7}
\end{figure}

The sensitivity of the output fluxes $F_\lambda$ on the abundances of 
H, C, N, O is of interest. First of all, the abundance changes 
affect the chemical balance, and hence the emitted spectra. Second, 
the structure of the model atmospheres is changed. This results in 
a complicated dependence of the spectral appearance on the abundances. 
Several model atmospheres were calculated for different abundances. 
A variation of the hydrogen abundance\footnote{In this work we will 
use an abundance scale $\sum N_i$ = 1.} from $\log\, N({\rm H}) = -4.0$ 
(hydrogen abundance in the R CrB atmosphere, see Asplund at al. 1997) 
to $\log N({\rm H})= -1.727$  (see ibid.) shows only a weak impact 
on the output spectrum of V4334 Sgr (Fig.~\ref{fig7}).

\begin{figure}
\resizebox{\hsize}{!}{\includegraphics{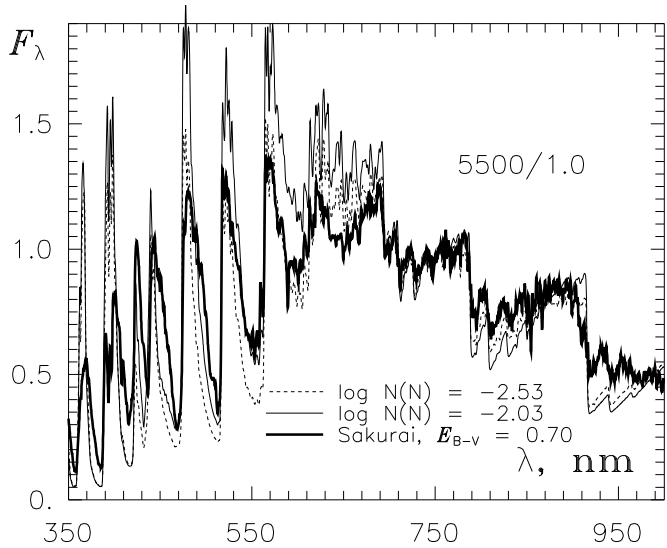}}
\caption{The dependence of theoretical spectra of V4334 Sgr on the
input nitrogen abundance.}
\label{fig8}
\end{figure}

As is obvious from Fig.~{\ref{fig3}}, changes of carbon, oxygen and 
nitrogen abundances affect the emitted spectra in different ways.
\CC bands lie in the blue, and CN bands lie in the red. The impact of 
changes of $\log\, N({\rm N})$ is shown in Fig.~{\ref{fig8}}.


\section{Discussion}

Our model computations yielded the following results:
\begin{itemize}
\item the overall shape of the SED of V4334 Sgr shows a critical dependence 
on \Tef{};
\item the comparison of observed and computed SEDs shows a reasonably 
good {\em qualitative\/} agreement. Thus, the contribution 
of various molecular opacity sources in the atmosphere of V4334 Sgr can be 
considered as well established;
\item at the same time, the comparison of observed and computed 
intensities of CN bands in the near UV part shows that the theoretical 
bands appear essentially deeper. This result may be interpreted as 
a lack of opacity in the blue part of the stellar spectrum 
(Yakovina \& Pavlenko 1998, Balachandran \& Bell 1998, Pavlenko \& 
Yakovina 1999) or as an impact of the dust (Duerbeck et al. 1999a).
\end{itemize}

No appropriate abundance determinations exist for April, 1997. Asplund 
et al. (1999) derived abundances for October 1996, which are probably best 
to be adopted in the absence of estimates for subsequent dates. A few
abundances were varied to study the impact of abundance changes on the 
emitted spectrum. As was shown in section 3.5, the overall shape of 
the SED mainly depends on \Tef{}. Therefore we do not expect our results 
to be crucially affected by abundances which differ somewhat from those 
found by Asplund et al. (1997). 

The weak dependence on hydrogen abundance $\log N ({\rm H})$ is especially 
noteworthy. Hotter models show a stronger dependence on hydrogen 
abundances (Asplund et al. 1997). In the case of lower \Tef, absorptions 
by atomic lines as well as molecular bands are still the main opacity 
sources. At \Tef{} = 5000\dots 6000 the maximum of the SED lies in a 
region governed by molecular bands. As a consequence, the dependence of 
model atmospheres and emitted SEDs on hydrogen abundance becomes weaker 
as compared to the case of higher \Tef{}. 

The ``missing opacity'' may be explained, at least partly, by the
incompleteness of the list of molecular bands used in this study. Due to 
the lack of input data, we do not take into account some molecular 
absorption bands at $\lambda < 400$ nm, like the \CC systems of 
Deslandres-d'Azambuja $C^1\Pi_g-A^1\Pi_u$ ($\lambda\lambda$ 339-378 nm), 
and $C^{'1}\Pi_g-A^1\Pi_u$ of Messerle-Krauss ($\lambda\lambda$ 339-378 nm).

Finally, several problems should be addressed briefly:

At the time when the 1997 spectrum was taken, no obvious dust obscuration
is seen in the light curve of V4334 Sgr. The first clear fading of the visual
light curve occured in February 1998, which was also accompanied by an 
abrupt reddening of colors (Liller et al. 1998, Duerbeck et al. 
1999a,b). Reports about infrared excesses and dust formation have been
made before (Duerbeck \& Benetti 1996, Kimeswenger et al. 1997,
Arkhipova et al. 1998 and Kamath \& Ashok 1999): most of these 
claim that dust had formed by March -- June 1997, but it appears that
it did not have a noticeable effect on magnitudes, colors, and thus
the SED in the optical region.

We carried out our \EBV{} determination as a self-consistent approach, 
i.e for the date of observations we obtained estimates of \Tef{}, 
$\log\, g$, and \EBV{}. If there was circumstellar extinction by dust, 
it would simply have increased the \EBV{}, since the reddening 
lines it the two-colour diagrams $(V-R)/(B-V)$ and $(V-I)/(B-V)$ 
are similar for interstellar reddening (Schultz \& Wiemer 1975), 
R CrB type circumstellar reddening (Lawson \& Cottrell 1989), and 
the reddening in V4334 Sgr in 1998 -- 1999 (Duerbeck et al. 1999b). 
If circumstellar reddening was present in V4334 Sgr in 1997, the value 
of \EBV{} = 0.7 would be an upper limit for the interstellar extinction.

In the case of the presence of some dust, the ``missing opacity'' 
in the blue part of the computed spectra (Figs.~\ref{fig5}-\ref{fig8}) 
may originate, at least partly, from scattering and/or absorption of 
dust particles, formed inside the atmosphere or in the envelope. 
The presence of dust inside and outside the atmosphere should affect its 
structure and spectrum in different ways. A dusty envelope cannot 
affect significantly the temperature structure of the photospheric 
layers. The envelope may produce a veiling of absorption lines, 
due to the formation of low-temperature blackbody radiation; but 
this effect should be noticeable only longwards of 1 $\mu$m, according 
to the models of Kipper (1999) for the spectrum of V4334 Sgr in 1997.

In the case of a dusty atmosphere, the interaction of grains
with the radiation field should change its internal structure, and 
therefore affect the spectrum. One may expect that the 
impact of the temperature change should be different for lines 
of atoms, ions and molecules. In order to consider the problem 
in the framework of a self-consistent approach, a more complicated 
model has to be used (Tsuji 1996). 

Our computations were carried out in the framework of the plane-parallel 
approach. We simplified the real situation (see Asplund et al. 1997) by 
ignoring sphericity effects. They should affect mainly the outermost 
layers, whose structure depends also on many other processes: 
formation of dust particles, depletion of molecular species, 
chromospheric-like effects, interaction with the dusty envelope, 
nonhomogeneity, etc. Thus, there are more problems to be studied 
even in the framework of this simple approach. Unfortunately, 
the physics of the processes mentioned above is poorly known.

On the other hand, direct computations kindly provided by T. Kipper 
(private communications) show that sphericity effects cannot be 
significant ($\Delta T < 120$ K in the outermost layers 
$\tau_{\rm ross} \sim 10^{-4}$) even for the 5500/0.0 model. The 
difference vanishes at $\tau_{\rm ross} \sim 10^{-1}$. At lower 
temperatures and higher luminosities, the sphericity effects are 
much larger.

For the model atmosphere 5500/1.0, an impact of sphericity effects 
appears to be negligible in view of the other uncertainties. 
Therefore, our determination of \Tef{} $\approx 5500$ K, $\log\, g 
\approx 0\dots 1$, \EBV{} $\approx 0.70$ for V4334 Sgr in April, 1997 
appears reliable.

In the frame of our paper we computed a grid of SEDs for model 
atmospheres of different \Tef. Differences in \Tef $\sim$ 250 K 
provide a noticeable effect, i.e. {\em formally\/} we may determine 
\Tef{} with an accuracy $\pm$ 200 K.

The true accuracy of the \Tef{} determination, given the uncertainties 
in the $\log\, g$, reddening and $\rm H+He$ abundances may be twice 
as large as the formal error (i.e. 400 K). On the other hand, our 
computations show a crucial dependence of the SED of V4334 Sgr 
on \Tef{}. Without doubt, this gives a good constraint for any physical 
models. Future studies should be accompanied by detailed photometric 
and spectroscopic determinations of abundances, gravities, etc. 

\begin{acknowledgements}
We thank Prof. T. Kipper for helpful discussions, the referee, 
Dr. M. Asplund, for helpful comments, and Dr. M. Turatto for taking 
the spectrum of V4334 Sgr.
\end{acknowledgements}

\begin{thebibliography}{}
\item[]{}Abia C., Pavlenko Y., de Laverny P., 1999, A\&A, 351, 273
\item[]{}Arkhipova V.P., Esipov V.F., Noskova R.I., Sokol G.V.,
         Tatarnikov A.M., Shenavrin V.I., Yudin B.F., Munari U., 
         Rejkuba M., 1998, Astr. Letters 24, 248
\item[]{}Asplund M., Gustafsson B., Lambert D.L., Rao N.K., 1997, 
         A\&A 321, L17
\item[]{}Asplund M., Lambert D.L., Kipper T., Pollacco D., Shetrone M.D., 
         1999, A\&A 343, 507
\item[]{}Balachandran S.C., Bell R.A. 1998, Nature 392, 791
\item[]{}Duerbeck H.W., Benetti S., 1996, ApJ 468, L111
\item[]{}Duerbeck H.W., Benetti S., Gautschy A., van Genderen A.M., 
         Kemper C., Liller W., Thomas T., 1997, AJ 114, 1657
\item[]{}Duerbeck H.W., van Genderen A., Jones A., Liller W., 1999a, 
         Southern Stars 38, 80
\item[]{}Duerbeck H.W., van Genderen A., Liller W., Sterken, C.,
         Benetti S., Arts J., Brogt E., Dijkstra R., Janson M., Kurk J.,
         van der Meer A., Voskes T., 1999b, submitted to AJ
\item[]{}Eyres S.P.S., Richards A.M.S., Evans A., Bode M.F., 1998, MNRAS
         297, 905
\item[]{}Gonzalez G., Lambert D.L., Wallerstein G., Rao N.K.,
         Smith V.V., McCarthy J.K., 1998, ApJS 114, 133
\item[]{}Hoffsaess D., 1979, Atomic Data and Nuclear Data Tables 24, 285. 
\item[]{}Fujimoto M.Y., 1977, PASJ 29, 331
\item[]{}Iben jr. I., Kaler J.B., Truran J.W., Renzini A., 1983, 
         ApJ 264, 605
\item[]{}Kamath U.S., Ashok N.M., 1999, MNRAS 302, 512
\item[]{}Kamenshikov V.A., Plastinin Yu.A, Nikolaev V.M., Novizkiy L.A., 
         1971, Radiative properties of gases for high temperatures, 
         Mashinostroenie, Moscow.
\item[]{}Kimeswenger S., Gratl H., Kerber F., Fouqu\'e P., Kohle S., Steele
        S., 1997, IAU Circ. 6608
\item[]{}Kimeswenger S., Kerber F., 1998, A\&A 330, L41
\item[]{}Kipper T., 1999, Inf. Bull. Var. Stars 4707
\item[]{}Kipper T., Klochkova V., 1997, A\&A 324, L65
\item[]{}Kurucz R., 1993, CD ROM No. 9
\item[]{}Lawson W.A., Cottrell P.L., 1989, MNRAS 240, 689
\item[]{}Liller W., Janson M., Duerbeck H., van Genderen, A., 1998, 
       IAU Circ. 6825 
\item[]{}Nakano S., Sakurai Y., et al., 1996, IAU Circ. 6322
\item[]{}Nersisyan S.E., Shavrina A.V., 
         Yaremchuk A.A., 1987, Astrophysics 30, 147.
\item[]{}Pavlenko Ya.V., 1997, Astrophys. Space Sci. 253, 43
\item[]{}Pavlenko Ya.V., 1999, Astr. Reports 43, 94
\item[]{}Pavlenko Ya.V., Yakovina L.A., 1999, Astr. Reports, accepted
\item[]{}Piskunov N.E., Kupka F., Ryabchikova T.A., Weiss W.W., 
         Jeffery C.S., 1995, A\&AS 112, 525
\item[]{}Pollacco D., 1999, MNRAS 304, 127
\item[]{}Schultz G.V., Wiemer W., 1975, A\&A 43, 133
\item[]{}Seaton M.J., Zeippen C.J., Tully J.A., et al., 1992, 
         Rev. Mexicana Astron. Astrophys. 23, 107.  
\item[]{}Tsuji T., 1973, A\&A 23, 411
\item[]{}Tsuji T., 1996, Dust formation in stellar photospheres.
         The case of carbon stars from dwarf to AGB, University of 
         Tokyo Preprint Series, 96-11, 1
\item[]{}Yakovina L.A., Pavlenko Ya.V., 1998, Kinematika Fiz. 
         Nebesn. Tel 14, 257
\end {thebibliography}
\end{document}